\documentclass[preprint,prb]{revtex4}
\usepackage{jpc}
\usepackage{epsfig}

\begin{document}

\title{Homogeneous Bubble Nucleation driven by local hot spots: a Molecular Dynamics Study}

\author{Zun-Jing Wang\footnote{electronic address: zunjingw@andrew.cmu.edu}}
\affiliation{  
FOM Institute for Atomic and Molecular Physics, Kruislaan 407, 1098 SJ Amsterdam, The Netherlands.\\
Department of Physics, Carnegie Mellon University, 5000 Forbes Avenue, Pittsburgh, PA 15213.
}

\author{Chantal Valeriani\footnote{electronic address: cvaleria@ph.ed.ac.uk}}
\affiliation{  
FOM Institute for Atomic and Molecular Physics, Kruislaan 407, 1098 SJ Amsterdam, The Netherlands.\\
School of Physics, James Clerk Maxwell Building, Kings Buildings, University of Edinburgh, Mayfield road, EH9 3JZ, Edinburgh, United Kingdom.
}

\author{Daan Frenkel\footnote{electronic address: df246@cam.ac.uk}}
\affiliation{  
FOM Institute for Atomic and Molecular Physics, Kruislaan 407, 1098 SJ Amsterdam, The Netherlands.\\
Department of Chemistry, University of Cambridge, Lensfield Road, CB2 1EW, Cambridge, United Kingdom.
}

\date{\today}

\vspace*{1cm}
\begin{abstract}

We report a Molecular Dynamics study of homogenous bubble
nucleation in a Lennard-Jones fluid. The rate of bubble nucleation
is estimated using forward-flux sampling  (FFS). We find that
cavitation starts with compact bubbles rather than with ramified
structures as had been suggested by Shen and Debenedetti ( {\it
J.~Chem.~Phys.~} {\bf 111},  3581 (1999)). Our estimate of the
bubble-nucleation rate is higher than predicted on the basis of
Classical Nucleation Theory (CNT). Our simulations show that local
temperature fluctuations correlate strongly with subsequent bubble
formation - this mechanism is not taken into account in CNT.
\end{abstract}


\maketitle


\section{Introduction}

Anyone who has ever sprinkled water droplets into a pan with hot
frying oil is familiar with the phenomenon of explosive boiling:
the liquid water will be heated far above its equilibrium boiling
point before the vapor phase nucleates suddenly and violently. As
this example illustrates, explosive boiling is very common, and
usually undesirable in practical situations. Bubble nucleation is
relevant in many different contexts: in addition to explosive
boiling~\cite{Shusser1999}, it plays a role in phenomena as
diverse as cavitation erosion~\cite{Brennen1995} and
sonochemistry~\cite{Suslick1990},  and it affects the design of
high-efficiency heat exchangers~\cite{Dhir1998}. In spite of the
practical relevance of these phenomena, the mechanism by which the
vapor phase nucleates from a homogeneous, super-heated liquid is
still under debate.

The standard theory used to describe the bubble nucleation
phenomenon is the so-called Classical Nucleation Theory (CNT). CNT
is commonly used to predict the rate of nucleation and estimate
the height of the free-energy barrier.~\cite{CNT1, CNT2, CNT3, CNT4, kelton} CNT
assumes that the growing droplet of the stable phase within the
metastable one is characterized by its bulk properties, it is
incompressible, and it has a spherical shape. Moreover, the
liquid-vapor interfacial free-energy does not depend on the
curvature of the droplet, condition that is satisfied at the
surface tension, that coincides with the equimolar dividing
surface in case of absence of  Gibbs absorption.

From an experimental point of view, CNT is in general used to
predict the nucleation rate, both in bubble nucleation and in
vapor condensation experiments. Early measurements of vapor
condensation performed by Blander and Katz~\cite{katz-bub} turned
out to be consistent with the CNT
predictions~\cite{Abraham1974,Frenkel1955,kelton}. However, more
recent measurements using a variety of different techniques and
materials~\cite{adams,Viisanen1994, hung, Wagner1984,Rudek1999,
Lihavainen2001, Hruby1996, Luijten1997, nozzle2004,nozzle2005,
nozzle2007,water,balibar1,balibar2,toluene,alco1,alco2} 
demonstrated that although CNT correctly
predicts the isothermal supersaturation-dependence of the
bubble-nucleation rate, it fails to predict the correct
temperature dependence: CNT predicts nucleation rates that are too
low at the low temperatures and too large at high temperatures
(even allowing for the fact that error bars may span several
orders of magnitude). To our knowledge, there exist no
experimental techniques that can directly probe the rate-limiting
step in the nucleation process, namely to formation of
``critical'' bubbles.

In an attempt to improve upon the predictions of CNT, Zeng and
Oxtoby~\cite{oxtoby-zeng} have used Density-Functional Theory (DFT) to
arrive at an estimate of bubble nucleation  rate. In particular,
Zeng and Oxtoby studied the liquid-to-vapor phase transition in a
Lennard-Jones fluid  and found that CNT underestimated the
nucleation rate by more than 15 orders of magnitude. In order to
explain the discrepancy between the DFT predictions and the CNT
estimates, Oxtoby and co-workers~\cite{oxtoby1, oxtoby2} pointed out that
CNT neglects all curvature corrections to the surface free-energy
of a droplet. Another curious feature of CNT that was already
pointed out by Cahn and Hilliard~\cite{cahn1,cahn2} and discussed
in some detail by Oxtoby and Evans~\cite{oxtoby3} is the fact that
this theory predicts a finite nucleation barrier at the point
where the metastable liquid undergoes spinodal decomposition $-$
this topic is still the subject of much debate.~\cite{Balsara1, Balsara2}
Delale et al.~\cite{delale} proposed a phenomenological estimate
of the minimum work of bubble formation, using experimental data
on super-heating. Using this approach, ref.~\onlinecite{delale}
estimated the steady-state bubble nucleation rate and found that,
as with the CNT estimates,  there was a large discrepancy between
the predicted computed and measured bubble-nucleation rates.

In 1999, Shen and Debenedetti~\cite{Shen1999} reported a Monte Carlo (MC) 
study to estimate the free-energy barrier for bubble
nucleation in a monatomic fluid of particles interacting via a
truncated Lennard-Jones potential. These authors computed the
free-energy barrier using Umbrella
Sampling~\cite{umbrella_torrie,umbrella_duijv}, and monitored the
progress of bubble formation using the total density of the system
as an order parameter. A geometric analysis of the resulting void
structures in the superheated liquid revealed that the formation
of the vapor phase proceeded via the formation of ramified,
system-spanning void structures. This observation disagrees
qualitatively with the predictions of CNT and DFT, both of which assume that the
critical nucleus is a compact, spherical bubble. It should however
be noted that ref.~\onlinecite{Shen1999} studied a system very close to
the point where the system is expected to undergo spinodal
decomposition.

The aim of the present paper is to study the kinetics of bubble 
nucleation under conditions similar to those studied in 
ref.~\onlinecite{Shen1999}. In order to study the time evolution 
of bubble nucleation, we must use Molecular Dynamics (MD) simulations 
and because it is then  better to use an interaction potential 
such that the forces  vanish continuously at the 
cut off, we could not use exactly the same model as Shen and Debenedetti.
Rather, we studied a system of particles interacting via a truncated, 
force-shifted Lennard-Jones potential. However, as we shall argue below, 
we study thermodynamic conditions that are comparable.
Using Forward-Flux Sampling (FFS)~\cite{FFS,FFS2},   we compute
the bubble nucleation rate and compare our numerical estimate with
the prediction of CNT. FFS is well suited
to simulate nucleation processes because it does not assume that
the nucleation process is slow compared to the time it takes all
other degrees of freedom to equilibrate. The latter assumption is
implicit in the Umbrella Sampling scheme.

The rest of this paper is organized as follows: we first describe
the simulations techniques used in our work,  then we present our
results and discuss the implications.
\section{Simulation Methods}

In order to probe the mechanism of bubble nucleation, we simulated
a system consisting of  3375 particles interacting via a truncated
and force-shifted (TSF) Lennard-Jones (LJ) pair potential:
\begin{equation}
v_{TSF}(r) = v_{LJ}(r) - v_{LJ}(r_{c}) - \left| \frac{d v_{LJ}}{dr} 
\right|_{r_{c}} (r-r_{c}) \text{   .}
\label{eq:potentialSF}
\end{equation}
The TSF-LJ potential is a modification of  the 12-6 Lennard-Jones
potential. The second and third terms on the right hand side of
Eq.~\ref{eq:potentialSF}  have been chosen such that both the
potential and its first derivative vanish continuously at
$r_{c}=2.5\sigma$ (where $\sigma$ is the particle diameter). The
reason why we use the TSF-LJ rather than the full LJ potential is
that during nucleation the system becomes inhomogeneous. Under
those conditions it is much simpler to use a finite-ranged
potential than an interaction with an $r^{-6}$-tail.

In what follows, we use reduced units: all physical variables are
expressed in terms of the Lennard-Jones units  $\sigma$, $m$,
$\varepsilon$, where $\sigma$ is a measure for the diameter of a
Lennard-Jones particle,  $m$ denotes its mass and $\varepsilon$ is
the depth of the attractive well of the (untruncated) LJ
potential.

We start our simulations by equilibrating the system in an $NPT$
ensemble by means of MD simulations with a
Nos\'{e}-Hoover thermostat~\cite{Nose1984,Hoover1985,frenkelsmit}
and an Andersen barostat~\cite{Andersen1980} (Appendix~A). In
order to study the bubble nucleation phenomenon, we use MD rather
than MC simulations, because standard MC algorithms
assume that the system is always locally in thermal equilibrium -
an assumption that may not be justified in the present case. In
all simulations, we adopt cubic periodic boundary conditions. The
equations of motion are integrated using a leap-frog
algorithm~\cite{Melchionna1993} with an integration time step of
$\Delta t^{\ast} =0.00046$, which corresponds to a time step of
about 1 femto-second (using the Lennard-Jones parameters for
Argon). In order to validate the MD code used in the present work,
we verified that it could successfully reproduce the liquid-vapor
coexistence curve of the TSF-LJ model that was computed by
Errington et al.~\cite{errington} (Appendix~B).

In an $NPT$ ensemble the simulation box adjusts itself in order to
keep both the temperature and the pressure of the system constant.
When a system undergoes a liquid-to-vapor during an $NPT$
simulation, the simulation box tends to expand dramatically: in
our simulation runs we limit such  ``explosions" by interrupting
the runs before the liquid phase has evaporated completely.

In order to study bubble nucleation, the system should be large
enough to accommodate a ``critical'' bubble (i.e. a bubble that is
equally likely to grow or shrink). On the basis of CNT we expect
that the size of the critical bubble is inversely proportional to
the degree of super-saturation.  If we carry out simulations at a
low supersaturation, the critical bubble size can become very
large and this would necessitate the use of very large simulation
boxes (containing a large number of particles). On the other hand,
if we choose the supersaturation too large, the liquid-vapor phase
transformation will not proceed via nucleation and growth but
through spinodal decomposition. These two constraints put rather
narrow limits on the values of supersaturation that can
conveniently be studied.

In order to obtain a rough estimate of the critical bubble size,
we make the CNT assumption that the chemical potential of the
vapor inside the critical bubble is the same as that of the liquid
outside. If we use the fact that the molar volume of the vapor
phase is much larger than that of the liquid phase, this implies
that pressure inside the critical bubble is the same as the
equilibrium vapor pressure at the imposed temperature.  The radius
of the critical bubble can then be approximated as:
\begin{equation} \label{rcrit}
r_{crit}=\frac{2\gamma}{P_{vp}(T)-P} \text{  ,}
\end{equation}
where $\gamma$ is the liquid-vapor surface tension, $P$ is the
imposed external pressure of the liquid and $P_{vp}(T)$ is the
equilibrium vapor pressure at the temperature $T$.

Shen and Debenedetti~\cite{Shen1999} studied bubble formation in
the  truncated Lennard-Jones system at a saturated vapor pressure
of $P^{\ast}=0.046$ and various degrees of super-heating, defined
as $S\equiv(T^{\ast}-T_{sat})/T_{sat}$, where $T_{sat}$ the
vapor-liquid coexistence temperature at the given pressure). For
$S$=9$\%$, CNT predicts a free-energy barrier of 37$k_{B}T$
for bubble nucleation.~\cite{Shen1999} In the present study, we do
not use exactly the same model as ref.~\onlinecite{Shen1999}, but we do
use the same degree of super-heating  ($S=9\%$). Although the
FTS-LJ model is not identical to the model used in
ref.~\onlinecite{Shen1999},  our simulations were carried out at
approximately the same distance from the critical point
($T^{\ast}/T^{\ast}_c$=0.914).  For this reduced temperature and
supersaturation, we expect that  a system of 3375 particles can
accommodate a critical bubble. To summarize: we carried out
simulations at a temperature $T^{\ast}=0.855$ and a reduced
pressure $P^{\ast}=0.026$. The coexistence temperature at this
reduced pressure is  $T^{\ast}_{sat}=0.785$. The  critical
temperature of the TSF-LJ model was estimated by Errington et
al.~\cite{errington} to be $T^{\ast}_c=0.935$. At coexistence
($T^{\ast}_{sat}=0.785$, $P^{\ast}_{sat}=0.026$), the
  number densities of the coexisting liquid and vapor are, respectively,
$\rho^{\ast}_{l, coex}=0.668$ and $\rho^{\ast}_{v, coex}=0.043$.
The number density of the superheated liquid at $T^{\ast}=0.855$
and $P^{\ast}=0.026$ is  $\rho^{\ast}_{l}=0.580$.

\subsection{Liquid-vapor surface tension \label{section:surfacetension}}
In order to compare our numerical estimate for the nucleation rate
with the prediction based on CNT, we need to know the value of the
surface tension $\gamma$. As  both the critical temperature and
the surface tension are sensitive to the long-range part of the
intermolecular potential,  it is important to have a reliable
estimate of $\gamma$ for the TSF-LJ potential truncated at
$r_c=2.5\sigma$ for the state point studied in the simulations.

To estimate $\gamma$, we have performed MD simulations of the
TSF-LJ liquid-vapor coexistent systems at $T^{\ast}=0.785$
($P^{\ast}=0.026$) and at $T^{\ast}=0.855$ ($P^{\ast}=0.046$)
separately.  We obtain the surface tensions of the planar
interfaces: $\gamma^{\ast}= 0.204 \pm 0.007$ at $T^{\ast}=0.785$,
and $\gamma^{\ast}= 0.098 \pm 0.008$ at $T^{\ast}=0.855$
(Appendix~C). Lutsko has used DFT 
to estimate the liquid-vapor surface tension of
the TSF-LJ system at the same state points and obtained
$\gamma^{\ast}_{DFT}=0.198$ at $T^{\ast}=0.785$, and
$\gamma^{\ast}_{DFT}=0.119$ at $T^{\ast}=0.855$.~\cite{lutsko2006,comlutsko}
These values for the surface tension are very similar to
the ones for the model studied in ref.~\onlinecite{Shen1999}.  In view of
the statistical errors in the simulation data, the agreement
between the DFT estimates for $\gamma^{\ast}$ and the simulation
results is quite good. We did not compute the surface tension
of a super-heated system at $T^{\ast}=0.855$ and $P^{\ast}=0.026$
because, under this thermodynamic condition a flat interface is
not thermodynamically stable.

\subsection{The order parameter \label{section:orderparameter}}
Even in the stable liquid phase, spontaneous density fluctuations
occur. In a superheated  liquid, some of these local density
fluctuations may grow to nucleate a bubble of the stable vapor
phase.  In order to follow the nucleation process we need an
``order parameter'' that provides a quantitative measure of the
degree of transformation from the liquid to the vapor phase. In
the present simulations we used  the volume of the biggest bubble
as an order parameter. This is a local order parameter that is
sensitive to the formation of large, connected void spaces. The
present definition of the order parameter is different from the
one used by Shen and Debenedetti, who used the  overall density of
the system as a global order parameter. We will return later to
this difference in the choice of order parameters.

We used the following procedure to identify the largest connected 
low-density region in the superheated liquid:
\begin{enumerate}
\item Following the method used in ref.~\onlinecite{Wang2005}, we construct
a three-dimensional grid with cell mesh of $(0.5\sigma)^3$  inside the 
simulation box ($\sigma$ is the LJ diameter).
\item As in ref.~\onlinecite{orderpar1}, we use Stillinger's cluster 
criterion~\cite{orderpar2} to identify the nearest neighbors of each 
particle: according to our definition, two particles are neighbors if 
their distance is less than a cut-off $r_{C}=1.6 \sigma$, corresponding to the 
first minimum of the radial distribution function of the liquid at the 
same thermodynamic conditions.
Using this criterion, we compute the probability distribution of
each particle's neighbors in both the liquid and the vapor phases.
\item We then identify a particle as  {\it liquid-like} if it has more 
than five neighbors. We used this particular threshold, as essentially all
particles in the homogeneous vapor phase have fewer neighbors and
almost all particles in a  homogeneous liquid phase have more (see
Figure~\ref{fig:disNN}).
\item We define a spherical volume with radius $r_C$ around every 
{\it liquid-like} particle:
all grid cells which are inside this spherical volume are labelled
as {\it liquid-like}. The remaining grid cells are labelled
{\it vapor-like}.
\item Finally, we perform a cluster-analysis on the {\it vapor-like} grid 
cells: a set of connected, {\it vapor-like} cells defines a bubble. We then 
select the largest of those and define this  as the largest bubble. The 
local order parameter is the  volume $W_{b}$ of the largest bubble 
expressed in units of $\sigma^3$ (for practical reasons,  we multiply 
the bubble volume by 10 and express it as an integer in FFS computations). 
\end{enumerate}

\subsection{Forward Flux Sampling with Molecular Dynamics}
FFS is a technique that allows us to track the pathways and flux
of a rare phase transition between two regions of the phase space,
the liquid ($L_{liq}$) and the vapor($V_{vap}$). As bubble
nucleation is a rare-event phenomenon, we incorporated 
FFS technique~\cite{FFS,FFS2}  in the MD
simulations of the super-heated fluid. There are specific
complications that arise when FFS is combined with MD and, to our
knowledge, this had not been attempted before.

Once we have defined a parameter $\lambda$  that measures the
progress of the nucleation process, we can define a value of
$\lambda$, denoted as $\lambda_0$, that define the boundary of the
metastable liquid region. By this we mean a value of $\lambda$
that is large enough to ensure that the overwhelming majority of
the states of the meta-stable liquid have a lower value of
$\lambda$, yet small enough to ensure that occasional a
spontaneous fluctuation will cross $\lambda_0$. We stress that the
choice of $\lambda_0$ is to some extent arbitrary and that our
results do not depend on the precise choice. Similarly, we define
a maximal value of $\lambda$, denoted as $\lambda_n$, such that
almost all trajectories that reach this value of $\lambda$ will
continue in the direction of progressive evaporation.  We denote
the domain of the meta-stable liquid ($\lambda<\lambda_0$) by
$L_{liq}$ and that of the vapor ($\lambda>\lambda_n$) by
$V_{vap}$.

As the transition from $L_{liq}$ to $V_{vap}$ is rare, the system
will initially spend most of its time in the $L_{liq}$-basin with
$\lambda<\lambda_0$. The key idea of FFS is to use a series of
non-intersecting hyper-surfaces (interfaces) in phase space, each
one identified by a value of the order parameter
$\lambda>\lambda_0$, to drive the system from $L_{liq}$ to
$V_{vap}$ in a ratchet-like manner.

The first step in the simulation is to compute $\Phi$, the flux of 
trajectories that cross $\lambda_0$ in the direction of increasing 
bubble size, per unit time and per unit volume. We store the 
configurations of the system at the time that a ``forward'' trajectory 
crosses $\lambda_0$. Once a set of such configurations have been 
collected at $\lambda_0$, we fire off stochastic trajectories that start 
from $\lambda_0$ and either reach the next interface $\lambda_1$ 
or fall all the way back to the initial basin $L_{liq}$ ($\lambda_0$). 
We then compute $P(\lambda_0|\lambda_1)$, the probability that
a trajectory that had crossed $\lambda_0$ continues to  $\lambda_1$, 
rather than return to the initial basin $L_{liq}$. We collect the configurations 
at $\lambda_1$ where trajectories from $\lambda_0$ first cross. Next we 
use these configurations to start new stochastic trajectories that 
will either reach $\lambda_2$ or fall back to $\lambda_0$.
In this way we compute $P(\lambda_1|\lambda_2)$, the
probability a trajectory started at $\lambda_1$ reaches $\lambda_2$ 
before returning to the initial basin $L_{liq}$. We then iterate this 
procedure all the way to the final boundary $\lambda_n$.
The FFS estimate for the nucleation rate is then expressed as the 
product of two terms:
\begin{equation}
R_{L_{liq}V_{vap}} = \Phi \cdot \prod_{i=0}^{n-1} 
P(\lambda_i|\lambda_{i+1}) \text{   .}
\label{eq:ffs1}
\end{equation}
The above description of the FFS technique makes it clear that
there is a problem when combining FFS with MD: FFS that the
underlying dynamics is stochastic, such that different
trajectories can be generated from the same initial point in phase
space. In contrast, standard Newtonian dynamics is deterministic,
hence there is only one trajectory emanating from a given point in
phase space. In order to combine FFS with MD
simulations in an $NPT$ {\it ensemble}, we make use of a weakly
stochastic variant of the (deterministic) Nos\'{e}-Hoover
thermostat. To this end, we add a Lowe-Andersen (LA) thermostat to the
Nos\'{e}-Hoover thermostat~\cite{Nikunen2003}. The LA thermostat 
is a momentum conserving and Galilean invariant
analog of the Andersen thermostat: it re-scales the relative
velocity between two particles from a Gaussian distribution with a
probability $\Delta t^{\ast} \nu^{\ast}$, where $\Delta t^{\ast}$
is the MD time step  and $\nu^{\ast}$ a tunable parameter
indicating the strength of the coupling between the system and the
thermostat (the smaller the $\nu^{\ast}$, the weaker the
coupling). Therefore the dynamical properties of the system can be
tuned by varying  $\nu^{\ast}$. The larger $\nu^{\ast}$ the more
stochastic the dynamics of our system. Of course, we wish to
ensure that the stochastic nature of the LA thermostat has only a
minor effect on the dynamical properties of the system. As a test,
we computed how the value of $\nu^{\ast}$ affectes the
self-diffusion coefficient in the FST-LJ system 
(see Figure~\ref{fig:diffuse}). We find that when
$\nu^{\ast} \leq 100$, the diffusion of the system with the LA
thermostat is indistinguishable from the one obtained with the
deterministic Nos\'{e}-Hoover  thermostat. In fact, there is
little change in the diffusion coefficient  as long as $\nu^{\ast}
\leq 1000$. However, when $\nu^{\ast} \geq 10000$, the LA
thermostat clearly affects the dynamics.

In order to balance the added stochasticity to the need
of keeping the system's dynamics as Newtonian as
possible, we choose $\nu^{\ast} = 500$.
As Figure~\ref{fig:divpath} shows, the added stochasticity is
enough to ensure divergence of the trajectories in phase 
space during the time it takes to travel from one interface to the next.~\cite{Bolhuis2003}


\section{Results and Discussion}

To perform an MD-FFS of bubble nucleation in a superheated liquid
system at $T^{\ast}=0.855$ and $P^{\ast}=0.026$, we locate the
boundary of the metastable liquid basin at $W_{b}=25$, and set the
interfaces at monotonically increasing values of the order
parameter $W_{b}$: $W_{b,1}=37$, $W_{b,2}=57$, $W_{b,3}=92$,
$W_{b,4}=120$, $W_{b,5}=180$, $W_{b,6}=240$, $W_{b,7}=300$,
$W_{b,8}=400$, $W_{b,9}=550$, $W_{b,10}=700$, $W_{b,11}=820$,
$W_{b,12}=950$, $W_{b,13}=1200$, $W_{b,14}=1900$. At each
interface we store $O(10^{2})$ configurations, select 30-50 of
these at random and fire  $O(10^{2})$ trajectories from each of
these. At a value of $W_{b} \sim 950$, approximately 50\% of all
trajectories continue to the vapor phase. This value of $W_{b}$
then constitutes our operational definition of the critical bubble
size. However, we stress that, as $W_{b}$ is not a perfect
reaction coordinate, the committor of individual configurations
with $W_{b}=950$ may differ from 50\%. We have not attempted to
optimize the choice of the reaction coordinate to make it coincide
more closely with an equi-committor surface. At $W_b=$3500, the
committor to reach the vapor phase is very nearly equal to one: we
use this value of $W_{b}$ as the boundary of the vapor phase. Once
simulations have reached this point, they are terminated.

Using FFS, we obtain the following estimate for the bubble nucleation rate
at $T^{\ast}=0.855$ and $P^{\ast}=0.026$: $ R=9 \times 10^{-15 \pm 1} 
\sigma^{-3} \tau^{-1}$. Using the Lennard-Jones parameters of Argon, 
this rate becomes $R=10^{19}$cm$^{-3}$s$^{-1}$. The computed 
bubble-nucleation rate is much higher than the one measured in early 
experiments for a number of volatile organic liquids: as an example, 
the bubble nucleation rate from ref.~\onlinecite{katz-bub} is $10^{4}-10^{6}$ 
bubbles cm$^{-3}$s$^{-1}$ at a temperature about $0.89T_c$ and an 
ambient pressure $1$ atm.

\subsection{Comparison with Classical Nucleation Theory}
A CNT-based theoretical estimate of the bubble nucleation rate was given 
by Katz  (Eq.~12 in Ref.~\onlinecite{katz-bub}):
\begin{equation}
\label{Katz}
R_{CNT}= N \left[ \frac{2\gamma}{\pi m B} \right]^{1/2} \exp (- \beta 
\Delta G)\text{   ,}
\end{equation}
where the prefactor ($N \left[ \frac{2\gamma}{\pi m B}
\right]^{1/2}$) contains the number density of the liquid ($N$)
and the Zeldovitch factor ($\left[ \frac{2\gamma}{\pi m B}
\right]^{1/2}$), $m$ is the mass of a molecule, $\gamma$ the
liquid-vapor surface tension, and $B$ a term that takes into
account the mechanical equilibrium of the bubble: in cavitation
experiments, $B$ is always equal to one. $\beta \Delta G$ is the
free-energy barrier to form the critical bubble, equal to
$\frac{16 \pi \gamma^{3} }{ 3 [P_V-P_L]^2}$ for a spherical
bubble.

The liquid-vapor surface tension of a planar interface of the
TSF-LJ fluid at $T^{\ast}=0.855$ is $\gamma^{\ast}= 0.098 \pm
0.008$ from  MD simulations. This leads to a free-energy
barrier of $\beta \Delta G= 47\pm 11$. Computing the bubble
nucleation rate by means of Eq.~\ref{Katz}, we find that CNT
predicts $R\approx 10^{-22}\sigma^{-3} \tau^{-1}$, with an
estimated error of five orders of magnitude in either direction.
The CNT prediction is six orders of magnitude less than the
simulation results. However, in view of the statistical errors in
the data, this discrepancy is only marginal. Using the DFT
estimate for the surface tension $\gamma^{\ast}= 0.119$ at
$T^{\ast}=0.855$, we obtain a CNT estimate $R\approx
O(10^{-36})\sigma^{-3} \tau^{-1}$, which is $21\pm 1$ orders of
magnitude smaller than the MD-FFS computation. Our finding is
consistent with earlier theoretical studies that found that CNT
estimates of bubble-nucleation rates were always lower than the
DFT results~\cite{oxtoby-zeng,oxtoby1,oxtoby2}. Of course, the DFT
estimate of the surface tension at $T^{\ast}=0.855$ refers to the
value for a planar interface at coexistence.

\subsection{Bubble nucleation pathways}

Figure~\ref{fig:pathsnap} shows a typical example of a computed pathway 
for bubble nucleation in a super-heated liquid at $T^{\ast}=0.855$ and $P^{\ast}=0.026$. 
As can be seen from Figure~\ref{fig:pathsnap}, the growing bubble is compact, 
even the critical bubble.   In this respect, our findings
appear to differ qualitatively from the findings of
ref.~\onlinecite{Shen1999}. As the present simulations were carried out
at virtually the same distance from the critical point and at the
same superheating, it is likely that the difference in the results
is due to a difference in the techniques that were used to follow
the progress of bubble nucleation. In particular, in the present
study we made use of a local order parameter (viz. the size of the
largest compact bubble), whereas ref.~\onlinecite{Shen1999} computed the
free-energy as a function of a global order parameter (viz. the
density of the entire system). These two methods need not probe
the same process. As discussed in ref.~\onlinecite{tenwolde} there are
two ways in which a system can accommodate an increase in a global
nucleation order parameter. The first is via the growth of a
compact nucleus, i.e. the process that we normally associate with
nucleation. The second is via the generation of many small nuclei
that, together, yield the same value of the global order
parameter. If the system is large enough and the surface tension
small enough, the second route will be favored for entropic
reasons.  As the global order parameter increases, the many small
nuclei may form a percolating structure, as was indeed  observed
in ref.~\onlinecite{Shen1999}.  The fact that in our FFS simulation we
do not observe this scenario but the nucleation of a compact
bubble indicates that the latter pathway is kinetically favored.

We stress that the use of an order parameter that is based on the
size of the largest bubble would still allow us to observe the
formation of ramified clusters if this were the favored route: it
would simply show up as sudden changes in the size of the largest
bubble as it merges with other (smaller) bubbles. Interestingly,
we occasionally do observe such processes, be it on a small scale:
we find that sometimes two small bubbles may merge to form a
larger bubble - we also observe the converse process where a
larger bubble spontaneously breaks up into two smaller bubbles.
When this happens the volume $W_b$ of the largest bubble changes
discontinuously (see Figure~\ref{fig:bubbleevolution1}). However,
unlike ref.~\onlinecite{Shen1999} we find that the resulting bubbles are
always compact.

The conclusion must therefore be that FFS would in principle allow
for the formation of ramified structures, as observed in
ref.~\onlinecite{Shen1999}. The fact that we do not see such structures
suggests that this pathway is kinetically unfavorable. Such
kinetic effects could not be observed in the equilibrium
umbrella-sampling study of ref.~\onlinecite{Shen1999}. There are,
however, other examples of nucleation events where the system
appears to prefer a higher free-energy route for purely kinetic
reasons.~\cite{Sanz2007}

\subsection{Local Kinetic Energy of Bubble Nucleus}
There exists and extensive literature on the study of rare events in 
condensed phases (see, for instance refs.~\onlinecite{Berne1997,Ferrario2006}).
In the numerical study of rare events, such as activated steps in 
chemical reactions or nucleation events, the reaction coordinate is 
usually constructed in configuration space. In other words: the most 
common situation is one where the success of a barrier-crossing 
trajectory does not depend on the initial momenta of the particles 
involved in the rare event. However, in bubble nucleation the situation 
may be different. To see this, consider the converse of bubble 
nucleation, viz. bubble collapse. Bubble collapse tends to proceed 
sufficiently rapidly that heat cannot be transported away quickly enough 
to keep the process at constant temperature. Hence, at the end of a 
bubble collapse the system may be locally very hot. This is the physical 
origin of sono-luminescence (see, e.g. ref.~\onlinecite{Brenner2002}).
It is therefore reasonable to assume that the reverse process, bubble 
nucleation, will preferentially originate in a part of the system that 
is locally hotter than its surroundings. Such an effect cannot be 
studied using a technique such as umbrella sampling, as this approach is 
based on the assumption of local thermal equilibrium. However, FFS makes 
no such assumptions and therefore allows us to see if local temperature 
fluctuations affect the propensity for bubble nucleation.

To investigate the relation between local heating and subsequent
nucleation, we studied the FFS trajectories of a large number of
successful bubble-nucleation events. We stress that the initial
conditions in these simulations were not biased: i.e. we did not
modify the initial temperature. Once a bubble had successfully
nucleated, we traced back the trajectory to the metastable liquid
basin and analyzed the local temperature along this trajectory. To
improve the statistics, we did this for a number of other
successful trajectories that started from the same initial
conditions. There are several ways to define the relevant local
temperature. We used the following procedure: we identified all
particles in the vapor phase inside the largest bubble at each
interface along FFS trajectory and then tracked the kinetic energy
of those particles from the basin to the end of FFS trajectories during
the bubble nucleation process. As an example, we show the ensemble 
of the configurations collected at each interface from the successful 
pathways at $T^\ast=0.855$ and $P^\ast=0.026$ 
($T^\ast_{sat}=0.785$) in Table~\ref{tavola}.

In order to see if temperature has an effect on the propensity for 
bubble nucleation we compared the local temperatures of bubbles that 
were committed to grow with that of all bubbles that have reached the 
same size. We can then compare these two temperatures with the 
temperature of the remainder of the system at the same point on the FFS 
trajectory. Figure~\ref{fig:ffsTdis} shows that the bulk liquid 
temperature of the total system computed with the $N_{i}^{s}$ collected 
MD snapshots at each interface $i$ is effectively constant and equal to 
the imposed temperature of thermostat.

Figure~\ref{fig:ffsTdis} shows that a successful nucleation is likely 
to have started in a region where the local temperature is higher than 
the temperature of the bulk liquid. In other words, a growing bubble is, 
at least initially, hotter than the surrounding liquid.
 From interface $6$, the local temperatures of the largest bubbles is 
not significantly different from the bulk temperature.
By analyzing the overall volume of the system, we find that it barely 
changes from interfaces $0\sim4$, whereas it starts increasing from 
interface $5$ (see Figure~\ref{fig:ffsVdis}). Interestingly, the system 
expansion starts at the point where the bubble has cooled to the 
temperature of the surroundings.

Between interface $5$ and the critical (iso-committor) surface $12$, 
there appears to be a difference between bubbles that continue to grow 
and those that shrink. On average, the temperature of growing bubbles 
appears lower than the ambient temperature, whilst is not significantly 
different from ambient. Note, however, the statistical errors are large: 
individual points may not conform to this general trend.  It should be 
noted that the effect of local temperature on nucleation may, in 
practice, be somewhat larger than indicated here because the system that 
we study is always weakly thermostatted. Hence, to return to the example 
of the collapsing bubble, the final temperature reached on collapse 
would be somewhat less in our simulations than in reality.

The main conclusion of the present section is that local
temperature fluctuation play an important role in bubble
nucleation. A description such as CNT that is based on the
assumption of local thermal equilibrium cannot capture this
phenomenon adequately. In the wider context of activated processes
in condensed media it is interesting to find an example of a rare
event where the kinetic energy appears to be a relevant quantity
to determine the propensity for the subsequent barrier crossing.
Such an observation would not be surprising for barrier crossings
that are largely ballistic, but it has, to our knowledge, not been
observed in the case of more diffusive barrier crossings. We
stress, however, that the effect of local temperature fluctuations
on bubble nucleation is not surprising: this phenomenon is
responsible for the operation of the bubble chamber to detect
elementary particles.


\section{Conclusions}
Combining Molecular Dynamics simulations with Forward-Flux
Sampling, we have studied the kinetics of bubble nucleation in a
truncated force-shifted Lennard-Jones liquid. In our simulations,
we used the volume of the largest bubble as the order parameter to
follow the phase transition. Using FFS, we computed the bubble
nucleation rate and compared it with the one predicted by CNT.
The simulated nucleation rates appear significantly larger
than those predicted by CNT. However, the CNT estimates are subject
to considerable uncertainty, as small statistical errors in the
calculated value of the surface tension have huge effects on the
estimates for the nucleation rate.  We note that the suggestion
that CNT underestimates bubble nucleation rates also follows from
a comparison of predictions based on CNT and
DFT~\cite{oxtoby-zeng}. Analysis of the nucleation pathway
indicates that nucleation takes place via the formation of compact
bubbles. These bubbles undergo shape fluctuations but are mostly
compact and fairly spherical. We do not find evidence for the
existence of ramified and percolating void
structures~\cite{Shen1999}. We argue that the difference between
our findings and those of ref.~\onlinecite{Shen1999} is not so much due
to a difference in the models studied but to the use of a global
order parameter in ref.~\onlinecite{Shen1999}. We find that local
temperature fluctuations are important for the propensity of
bubble nucleation. At the beginning of nucleation event, the
incipient bubble is always significantly hotter than the
surrounding bulk liquid. When the total volume of the system starts 
increasing,  the bubbles cools down to ambient temperature.

\bigskip

\begin{acknowledgments}
The authors thank J.~R.~Errington for sharing his data of the
truncated force-shifted phase diagram, J. ~Lutsko for providing us
the DFT estimate of the liquid-vapor surface tension, and
J.~van Meel for a careful reading of the manuscript. The work of
the FOM Institute is part  of the research program of the
Stichting voor Fundamenteel Onderzoek der Materie (FOM), which is
financially supported by the Nederlandse Organisatie voor
Wetenschappelijk Onderzoek (NWO).
\end{acknowledgments}

\clearpage
\section*{Appendix A: Details on the used thermostat and barostat \label{app:termbar}}
\addcontentsline{toc}{section}{Appendix A: Details on the used thermostat and barostat \label{app:termbar}}
\renewcommand{\theequation}{A-\arabic{equation}} 
\setcounter{equation}{0} 

In our simulations, we use a thermostat and a barostat to keep the thermodynamic conditions of the system constant. However, we wish to work under conditions where the temperature and pressure control do not affect the dynamics of the system appreciably on the timescale for molecular velocity fluctuations.
To select reasonable relaxation times for both the thermostat and the barostat, we performed several trial runs at different temperatures and pressures before selecting $\tau_T^{\ast}=0.093$ for the thermostat and $\tau_P^{\ast}=1.390$ for the barostat. Figure~\ref{fig:tempPdis} illustrates the (kinetic) temperature fluctuations in the fluid about a value of $\langle T^{\ast}\rangle=0.900$ and the (virial) pressure fluctuations in the same fluid about a value of $\langle P^{\ast}\rangle=0.064$.

The mass of the  ``piston'' in the barostat was chosen such that the timescale  for volume fluctuations corresponded to the time it takes a sound wave to traverse the simulation box (approximately $1$ ps).~\cite{Andersen1980}

\section*{Appendix B: TSF-LJ liquid-vapor phase diagram \label{app:phasediag}}
\addcontentsline{toc}{section}{Appendix B: TSF-LJ liquid-vapor
phase diagram \label{app:phasediag}}
\renewcommand{\theequation}{B-\arabic{equation}} 
\setcounter{equation}{0} 

Errington et al.~\cite{errington} computed the liquid-vapor phase
diagram of the TSF-LJ by means of MC simulations. In
fig.~\ref{fig:phasediagrame1} we show Errington's data for the
TSF-Lennard Jones ($r_{c}=2.5\sigma$). For the sake of comparison,
we also reproduce the liquid-vapor phase diagram data of the full
Lennard Jones calculated by Potoff et al.~\cite{critpanaio} and by
Smit~\cite{critsmit} and the corresponding results for the
truncated and shifted Lennard-Jones at $r_{c}=2.5\sigma$
calculated by Smit~\cite{critsmit} and by Trokhymchuk et
al.~\cite{alejandre}. In the same figure, we also plot our own
simulation data for the TSF-LJ potential. We find excellent
agreement between our own simulation results and the data of
ref.~\onlinecite{errington}.

As was already pointed out by Smit~\cite{critsmit}, truncating
and shifting the potential has a large effect on the location of the critical temperature.
Figure~\ref{fig:phasediagrame1}  illustrates this observation.

\section*{Appendix C: liquid-vapor surface tension computed with MD simulations \label{app:ST-MD}}
\addcontentsline{toc}{section}{Appendix C:  liquid-vapor surface
tension computed with MD simulations \label{app:ST-MD}}
\renewcommand{\theequation}{C-\arabic{equation}} 
\setcounter{equation}{0} 

To compute the liquid-vapor surface tension, we start by preparing
a half-liquid-half-vapor system with 16000 particles. After having
equilibrated a liquid by means of an NPT simulation at the
coexistence temperature and pressure ($T^{\ast}=0.785$,
$P^{\ast}=0.026$ or $T^{\ast}=0.855$, $P^{\ast}=0.046$) , we
prepare a liquid-vapor interface by evaporating half of the
system. The thickness of the liquid slab and the size of
simulation box are chosen sufficiently large to make finite-size
effects unlikely. Once the system of coexisting liquid and vapor
has reached thermodynamic equilibrium, we calculate the
liquid-vapor surface tension in an NVT ensemble (see
ref.~\onlinecite{wang2001}). We compute the density profile $\rho(z)$,
and the normal and tangential pressure profiles, $P_{N}(z)$ and
$P_{T}(z)$ along the $z$ direction perpendicular with respect to
the liquid-vapor dividing interface (see Figure~\ref{fig:profilerhoP}),
and use the Irving-Kirkwood approach to compute the pressure
tensor~\cite{Walton1983}. The normal component of the pressure
tensor is expressed as
\begin{equation}
P_{N}(z) =\left<\rho(z)\right>k_{B}T- \frac{1}{A}\left< \sum_{i}\sum_{j>i} \frac{z_{ij}^2}{r_{ij}} \frac{dU(r_{ij})}{dr_{ij}}\frac{1}{\left|z_{ij}\right|}\times\theta\left(\frac{z-z_{i}}{z_{j}-z_{i}}\right)\theta\left(\frac{z_{j}-z}{z_{j}-z_{i}}\right)\right> \text{   ,}
\label{eq:surfacePN}
\end{equation}
whereas the tangential component is given by
\begin{equation}
P_{T}(z) =\left<\rho(z)\right>k_{B}T- \frac{1}{A}\left< \sum_{i}\sum_{j>i} \frac{x_{ij}^2+y_{ij}^2}{2r_{ij}} \frac{dU(r_{ij})}{dr_{ij}}\frac{1}{\left|z_{ij}\right|}\times\theta\left(\frac{z-z_{i}}{z_{j}-z_{i}}\right)\theta\left(\frac{z_{j}-z}{z_{j}-z_{i}}\right)\right> \text{   ,}
\label{eq:surfacePT}
\end{equation}
where $\rho(z)$ denotes the density profile along the z direction, $k_{B}$ is the Boltzmann constant,
$T$ the absolute temperature, $A=L_x\times L_y$ the area of the simulation box in the $x$-$y$ plane,  $r_{ij}$ the distance between two particles $i$ and $j$, and $U$ the TSF-LJ internal energy (with cut-off $r_c^{\ast}=2.5\sigma$). The angular brackets denote a canonical average.
For computational purposes, we slice the simulation box into a
large number of   slabs in the $z$ direction, where the width of
each slab is set to $\Delta z \approx 0.1$ $\sigma$ (corresponding
to 1141 slabs at $T^{\ast}=0.785$ and 1160 slabs at
$T^{\ast}=0.855$ respectively). The surface tension is given
by
\begin{equation}
\gamma = \frac{1}{2}\int_{0}^{L_z}\left(P_N(z)-P_T(z)\right)dz  \text{  .}
\label{eq:surfacetension}
\end{equation}
The surface tensions that we obtain from MD simulations  are
$\gamma^{\ast}= 0.204 \pm 0.007$ at $T^{\ast}=0.785$,
$P^{\ast}=0.026$ and $\gamma^{\ast}= 0.098 \pm 0.008$ at
$T^{\ast}=0.855$, $P^{\ast}=0.046$.



\newpage
\centerline{\bf TABLES}

\vspace*{2cm}

\begin{table}[h!]
\caption{\label{tavola} Statistics collected from different
sets of FFS ``bubble-nucleation'' trajectories at $T^\ast=0.855$
and $P^\ast=0.026$. $(W_{b})_i$ denotes the order parameter at
each interface $i$ as defined in section
\ref{section:orderparameter}. $N_{i}^{s}$ is the total number of
MD configurations (including  momenta) collected at each interface
$i$, irrespective of whether the bubble will continue to grow or
it will collapse after interface $i$. $N_{i}^{p}$ is the average
number of  particles in the vapor phase inside or near the
largest bubble. In this set, we include all particles that are
inside or within $2.5\sigma$ from the large bubble, identified
using the approach described in section~\ref{section:orderparameter}. 
From interface $14$, bubble nearly always grows without collapse.}
\vspace{1cm}
\centerline{
\begin{tabular}{|cccc|cccc|cccc|cccc|cccc|cccc|}
\hline \hline 
$i$\,\, &$(W_{b})_i$ &$N_{i}^{s}$ &$N_{i}^{p}$ \,\,&\,\,$i$\,\,&$(W_{b})_i$ &$N_{i}^{s}$ &$N_{i}^{p}$\,\,&\,\,$i$\,\,&$(W_{b})_i$ &$N_{i}^{s}$ &$N_{i}^{p}$\,\,&\,\,$i$\,\,&$(W_{b})_i$ &$N_{i}^{s}$  &$N_{i}^{p}$\,\,&\,\,$i$\,\,&$(W_{b})_i$ &$N_{i}^{s}$  &$N_{i}^{p}$ \\
\hline \hline 
0  &25   &4012  &50	&3  &92  &1795	 &79 &6  &240  &725 &121 &9 &550  &334 &188  &12 &950 &79 &264  \\
\hline  
1  &37   &3619  &60	&4  &120  &548	 &98 &7  &300   &285 &142 &10 &700  &329  &203 &13 &1200 &147 &271 \\ 
\hline  
2  &57   &1304  &70	&5  &180  &512  &103    &8  &400  &320  &170   &11 &820   &374  &215  &14 &1900 &174  &289 \\
\hline \hline 
\end{tabular}}
\label{table:ffsensemble}
\end{table}


\newpage
\centerline{\bf FIGURE CAPTIONS}

\begin{figure}[pht!]
\caption{Probability distribution of the number of neighbors $N$ for both the liquid and the vapor at coexistence $T^\ast=0.785$ and $P^\ast=0.026$. Two particles are neighbors if their distance is smaller than $r_C=1.6 \sigma$.}
\label{fig:disNN} 
\end{figure}

\begin{figure}[pht!]
\caption{Diffusion coefficients as a function of time in the MD simulations with a Lowe-Andersen thermostat.  The effect of $\nu^{\ast}$ on self-diffusion coefficient of the TSF-LJ fluid is shown. }
\label{fig:diffuse} 
\end{figure}

\begin{figure}[pht!]
\caption{MD simulation runs from the first to the second interface ($\nu^{\ast} = 500$) in a TSF-LJ system at $T^{\ast}=0.855$ and $P^{\ast}=0.026$.}
\label{fig:divpath}
\end{figure}

\begin{figure}[pht!]
\caption{Typical bubble-nucleation pathway in a super-heated liquid at $T^{\ast}=0.855$ and $P^{\ast}=0.026$. The top left frame shows the initial stage of the bubble, while the bottom right snapshot is the critical bubble.}
\label{fig:pathsnap}
\end{figure}

\begin{figure}[pht!]
\caption{Fluctuations of the order parameter as a function of MD time steps for a generic FFS-interface at $T^{\ast}=0.855$ and $P^{\ast}=0.026$. The jumps in $W_b$ are due to the merge and break-up of smaller bubbles. }
\label{fig:bubbleevolution1} 
\end{figure}

\begin{figure}[pht!]
\caption{Temperature at each interface from MD-FFS at $T^{\ast}=0.855$ and $P^{\ast}=0.026$. The green line denotes the temperature of the thermostat used in MD. The entire liquid bulk temperature is denoted with black solid square. The local temperature of the  ``both-backward-and-forward" largest bubble (blue open triangle) and the  ``only-forward" largest bubble (red solid circle) are also shown.}
\label{fig:ffsTdis}
\end{figure}

\begin{figure}[pht!]
\caption{System volume at each interface from MD-FFS at $T^{\ast}=0.855$ and $P^{\ast}=0.026$. An expansion of the overall volume of the system is observable from interface $5$.}
\label{fig:ffsVdis}
\end{figure}

\renewcommand{\thefigure}{A-\arabic{figure}}
\setcounter{figure}{0} 

\begin{figure}[pht!]
\caption{Comparison of temperature (left panel) and pressure (right panel) distributions between MD simulations and Gaussian distributions peaked at the same average value (continuous curves).}
\label{fig:tempPdis}
\end{figure}

\renewcommand{\thefigure}{B-\arabic{figure}}
\setcounter{figure}{0} 

\begin{figure}[pht!]
\caption{$T-\rho$ liquid-vapor phase diagram. The filled squares denote
the present simulation results of the TSF-LJ fluid ($r_{c}=2.5\sigma$).  
Errington's results for the same system are shown as a drawn curve. 
All other (open) symbols refer to simulation data for other variants of the Lennard-Jones model
(see text).}
\label{fig:phasediagrame1}
\end{figure}

\renewcommand{\thefigure}{C-\arabic{figure}}
\setcounter{figure}{0} 

\begin{figure}[pht!]
\caption{Profiles of $\rho^{\ast}(z)$ (top panel), $P^{\ast}_N(z)$ and
$P^{\ast}_T(z)$ (bottom panel, black-solid and red-dashed lines,
respectively) of the liquid-vapor coexisting system at $T^{\ast}=0.785$ 
from MD simulations. We verified that the number densities of the
liquid and vapor are the same as those of a saturated liquid and
vapor at imposed temperature  ($T^{\ast}=0.785$), and that the
normal pressure equals the saturated vapor pressure
$P^{\ast}=0.026$ at this temperature.}
\label{fig:profilerhoP}
\end{figure}

\newpage
\clearpage
\centerline{\bf FIGURES}
\vspace*{2cm}

\begin{figure}[h]
\epsfxsize=12.0cm
\epsfbox{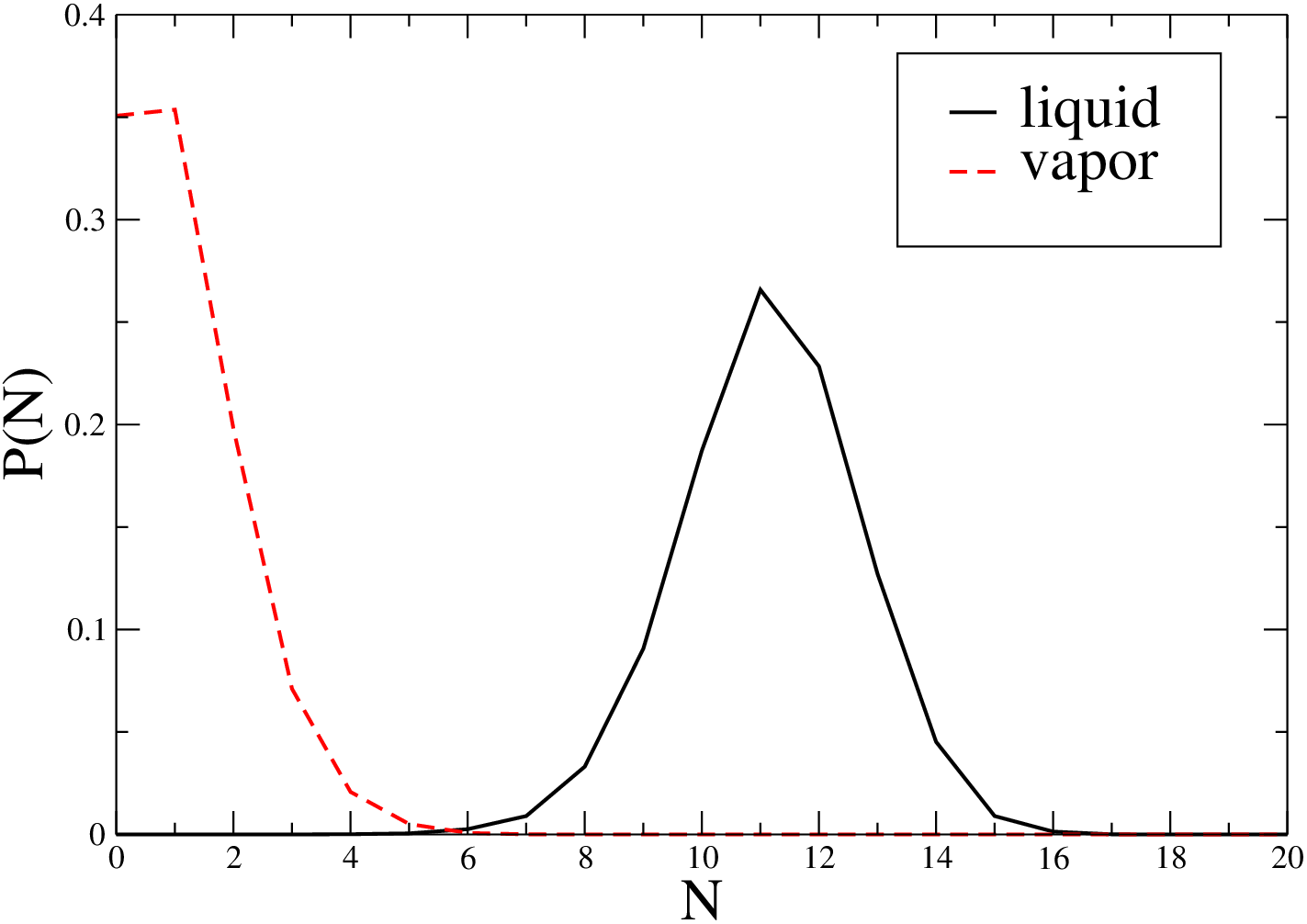}
\end{figure}
\vspace*{1.0cm}
Figure \ref{fig:disNN}

\newpage
\vspace*{2cm}

\begin{figure}[h]
\epsfxsize=12.0cm
\epsfbox{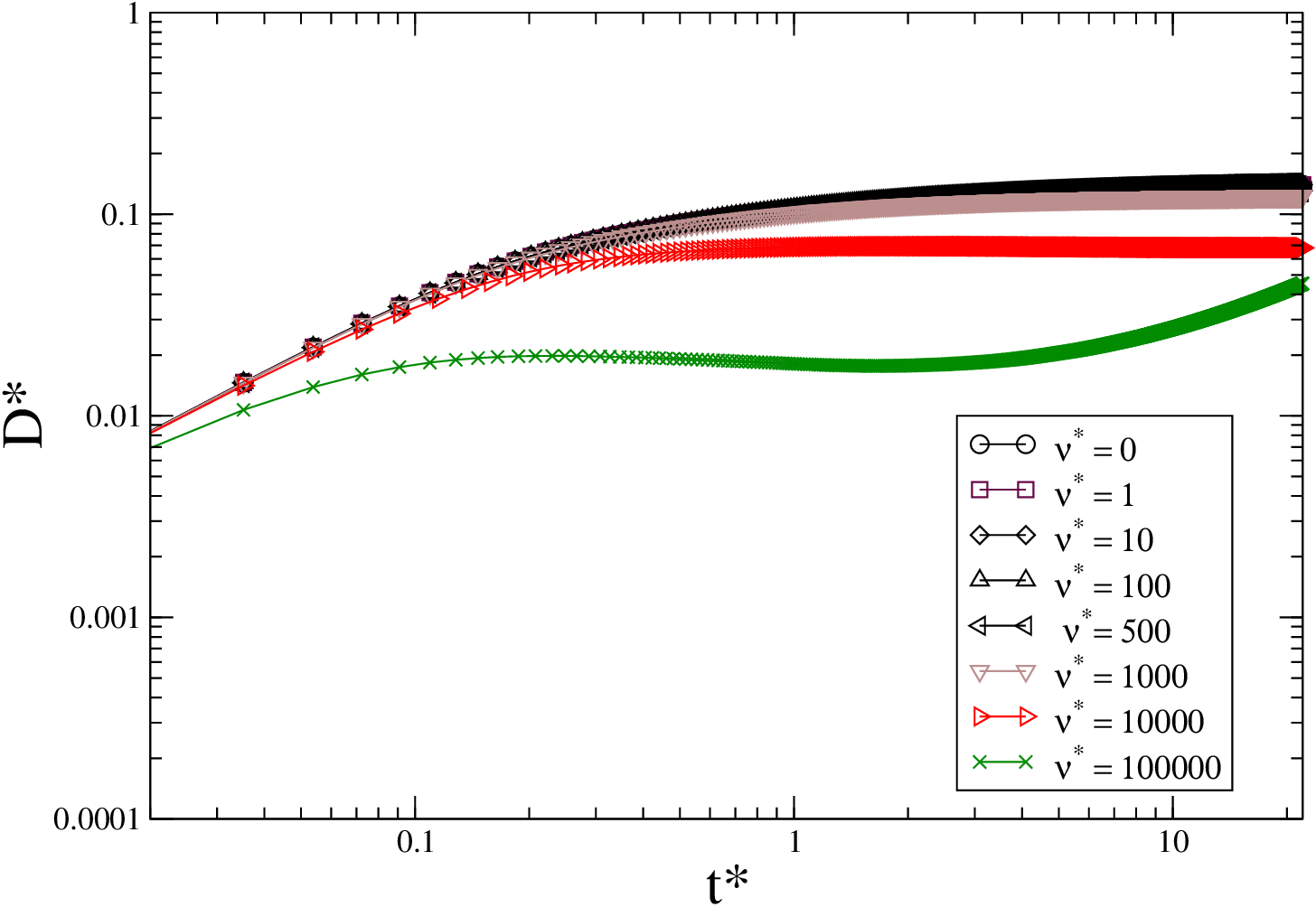}
\end{figure}
\vspace*{1.0cm}
Figure \ref{fig:diffuse}

\newpage
\vspace*{2cm}

\begin{figure}[h]
\epsfxsize=12.0cm
\epsfbox{fig3.eps}
\end{figure}
\vspace*{1.0cm}
Figure \ref{fig:divpath}

\newpage
\vspace*{2cm}

\begin{figure}[h]
\epsfxsize=3.5truecm
\epsfbox{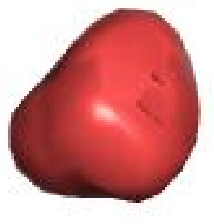}
\hspace*{0.5cm}
\epsfxsize=3.5truecm
\epsfbox{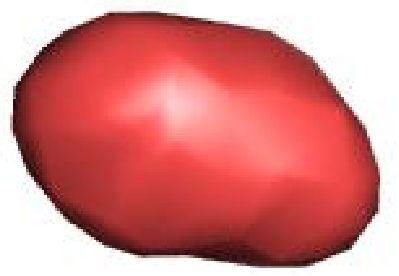}
\hspace*{0.5cm}
\epsfxsize=3.5truecm
\epsfbox{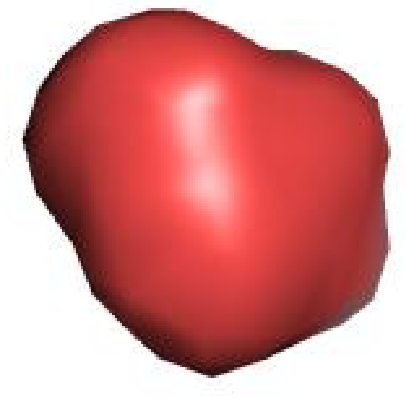}
\vspace*{1.0cm}

\epsfxsize=3.5truecm
\epsfbox{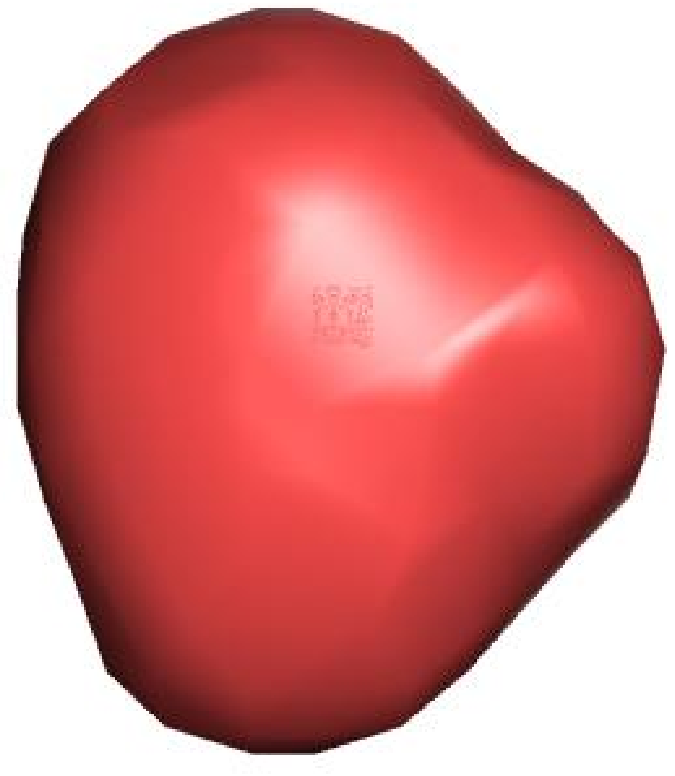}
\hspace*{0.5cm}
\epsfxsize=3.5truecm
\epsfbox{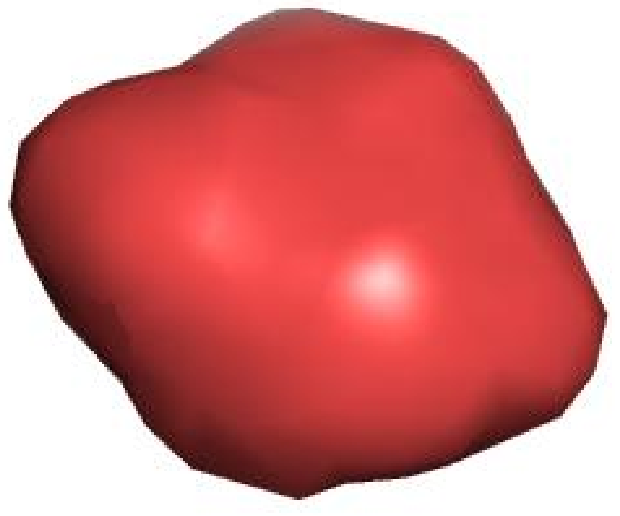}
\hspace*{0.5cm}
\epsfxsize=3.5truecm
\epsfbox{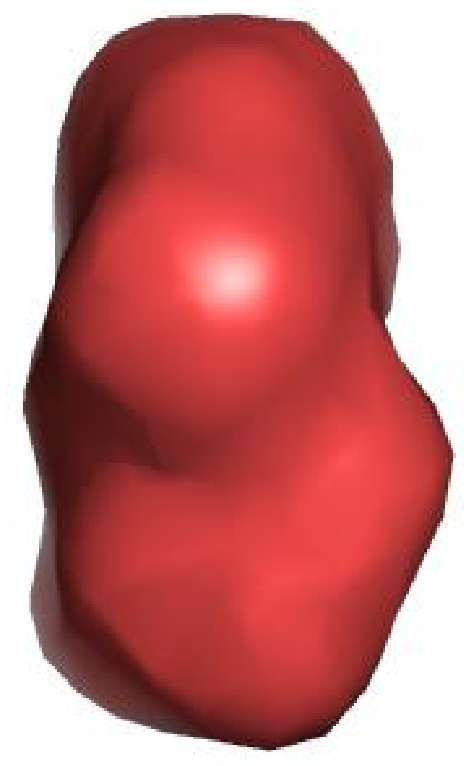}
\vspace*{1.0cm}

\epsfxsize=3.5truecm
\epsfbox{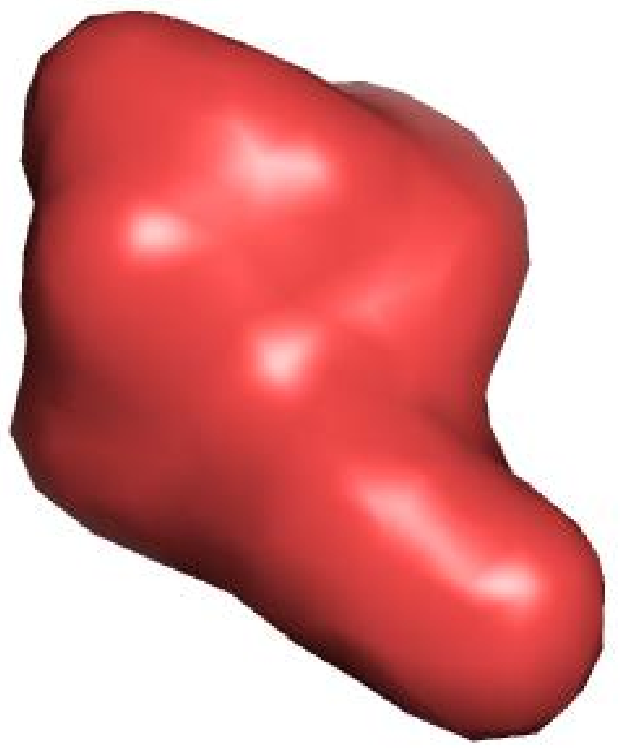}
\hspace*{0.5cm}
\epsfxsize=3.5truecm
\epsfbox{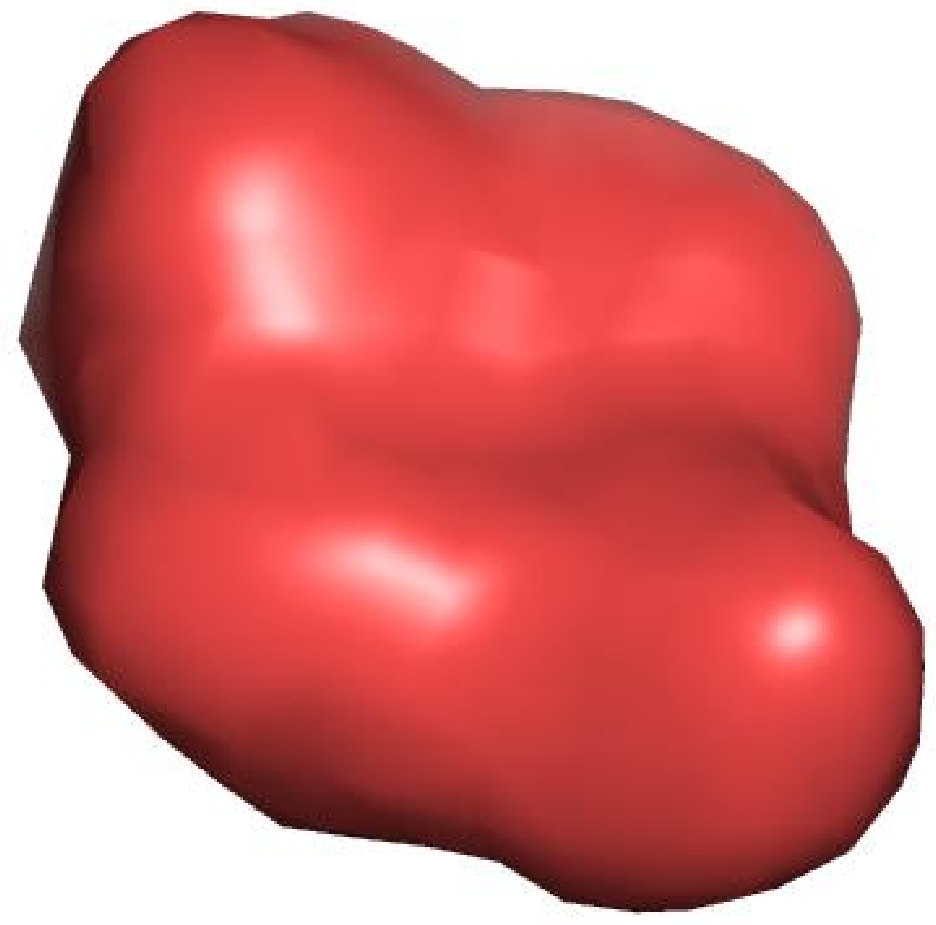}
\hspace*{0.5cm}
\epsfxsize=3.5truecm
\epsfbox{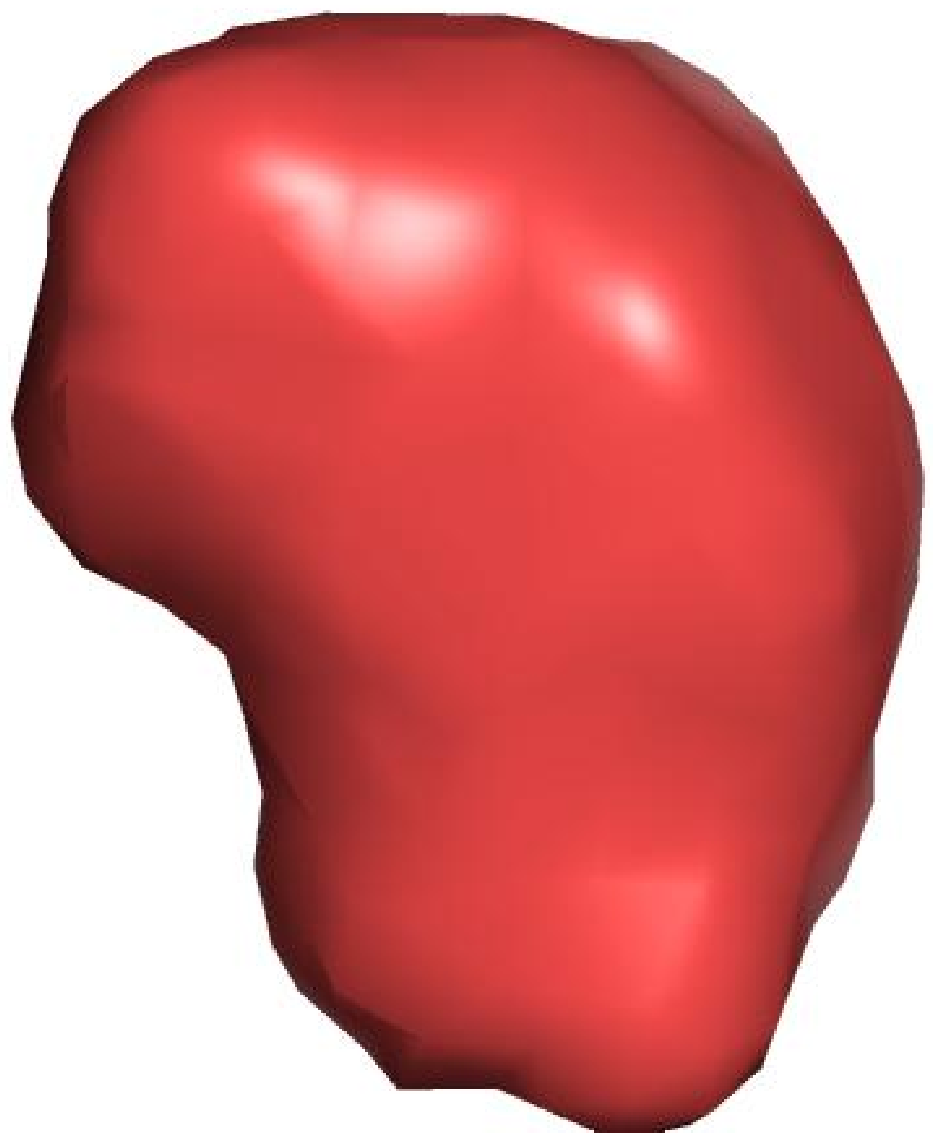}
\vspace*{2cm}
\end{figure}
Figure \ref{fig:pathsnap}

\newpage
\vspace*{2cm}

\begin{figure}[h]
\epsfxsize=12.0cm
\epsfbox{fig5.eps}
\end{figure}
\vspace*{1.0cm}
Figure \ref{fig:bubbleevolution1}

\newpage
\vspace*{2cm}

\begin{figure}[h]
\epsfxsize=12.0cm
\epsfbox{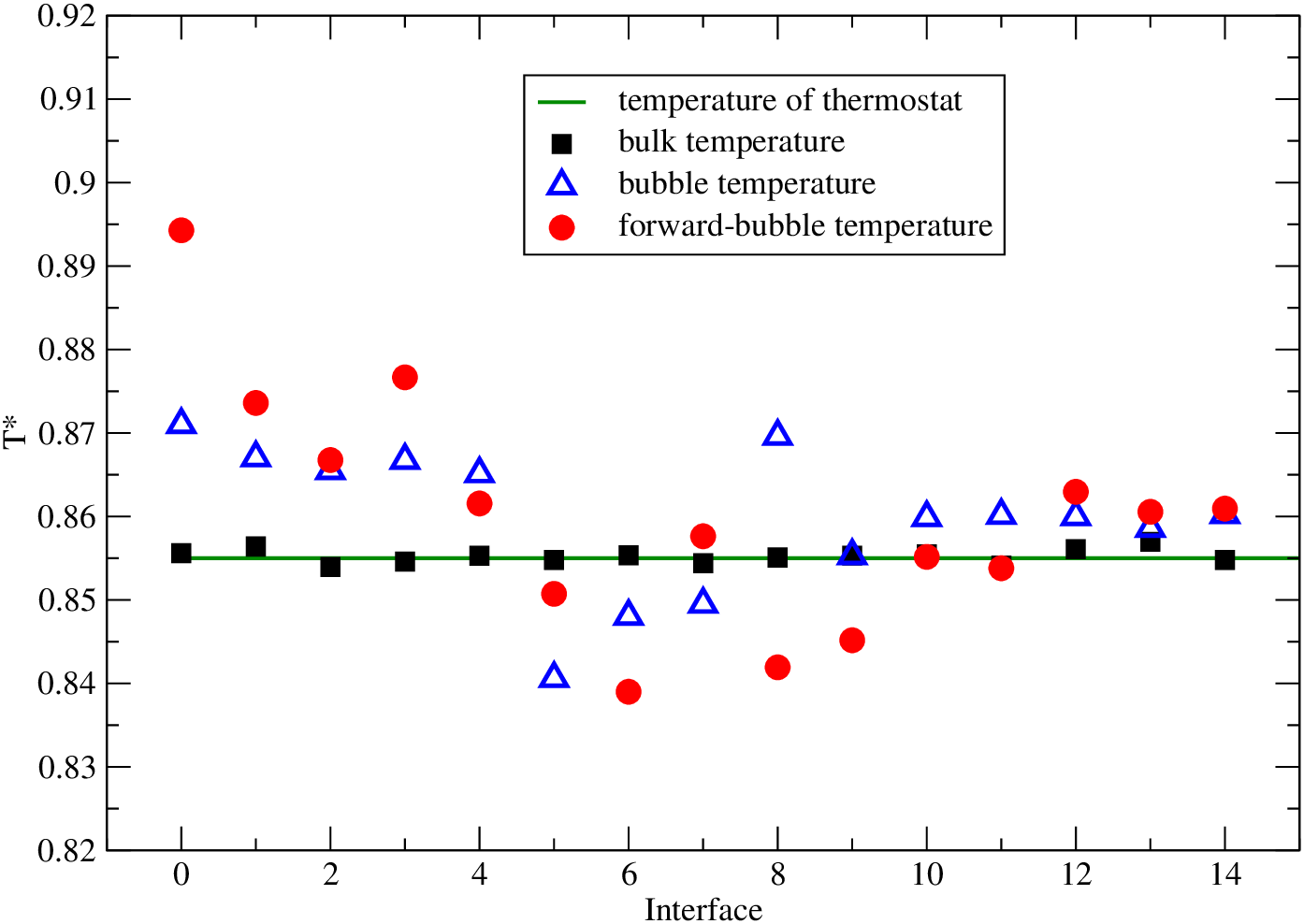}
\end{figure}
\vspace*{1.0cm}
Figure \ref{fig:ffsTdis}

\newpage
\vspace*{2cm}

\begin{figure}[h]
\epsfxsize=12.0cm
\epsfbox{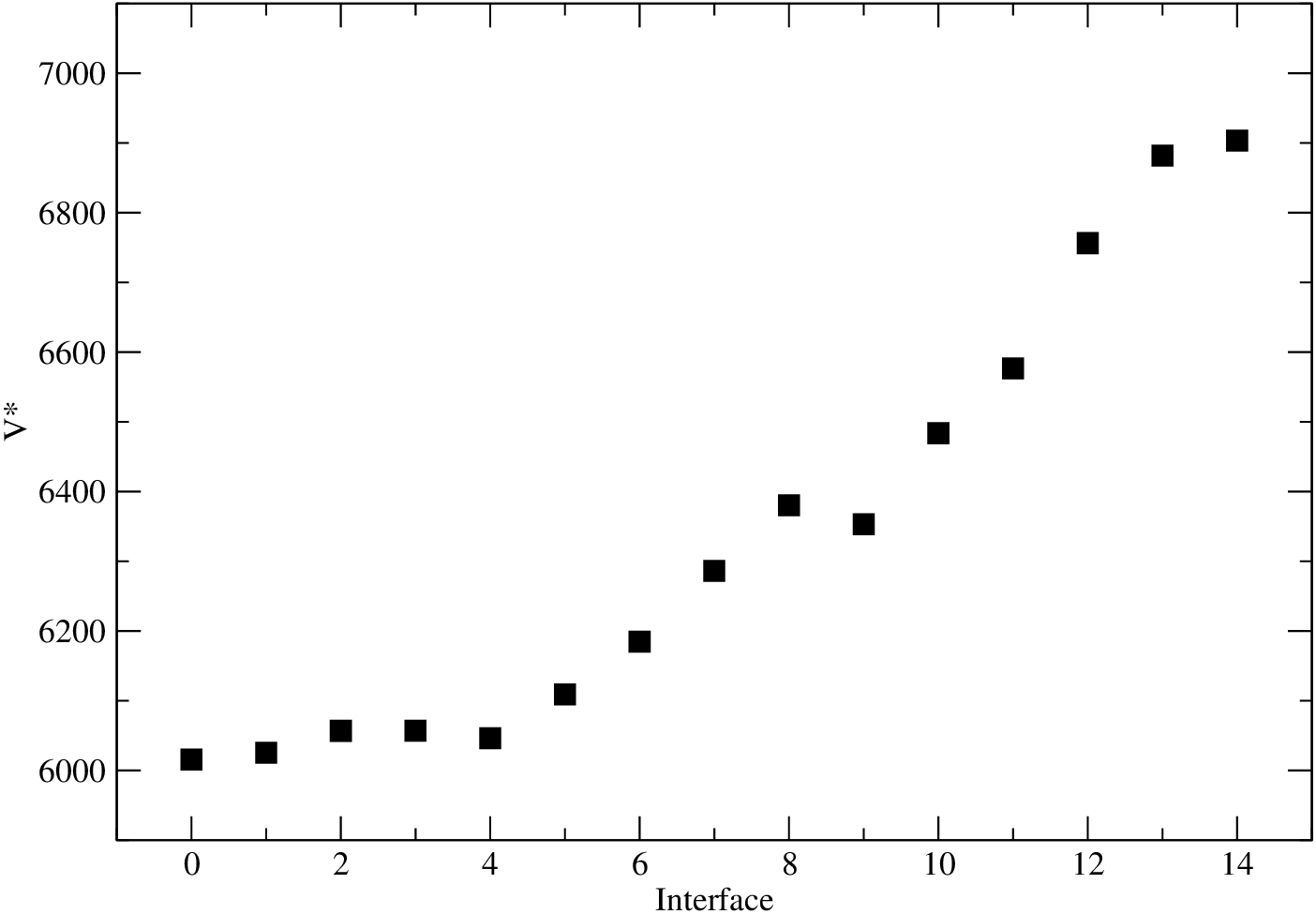}
\end{figure}
\vspace*{1.0cm}
Figure \ref{fig:ffsVdis}

\newpage
\vspace*{2cm}

\begin{figure}[h]
\epsfxsize=12.0cm
\epsfbox{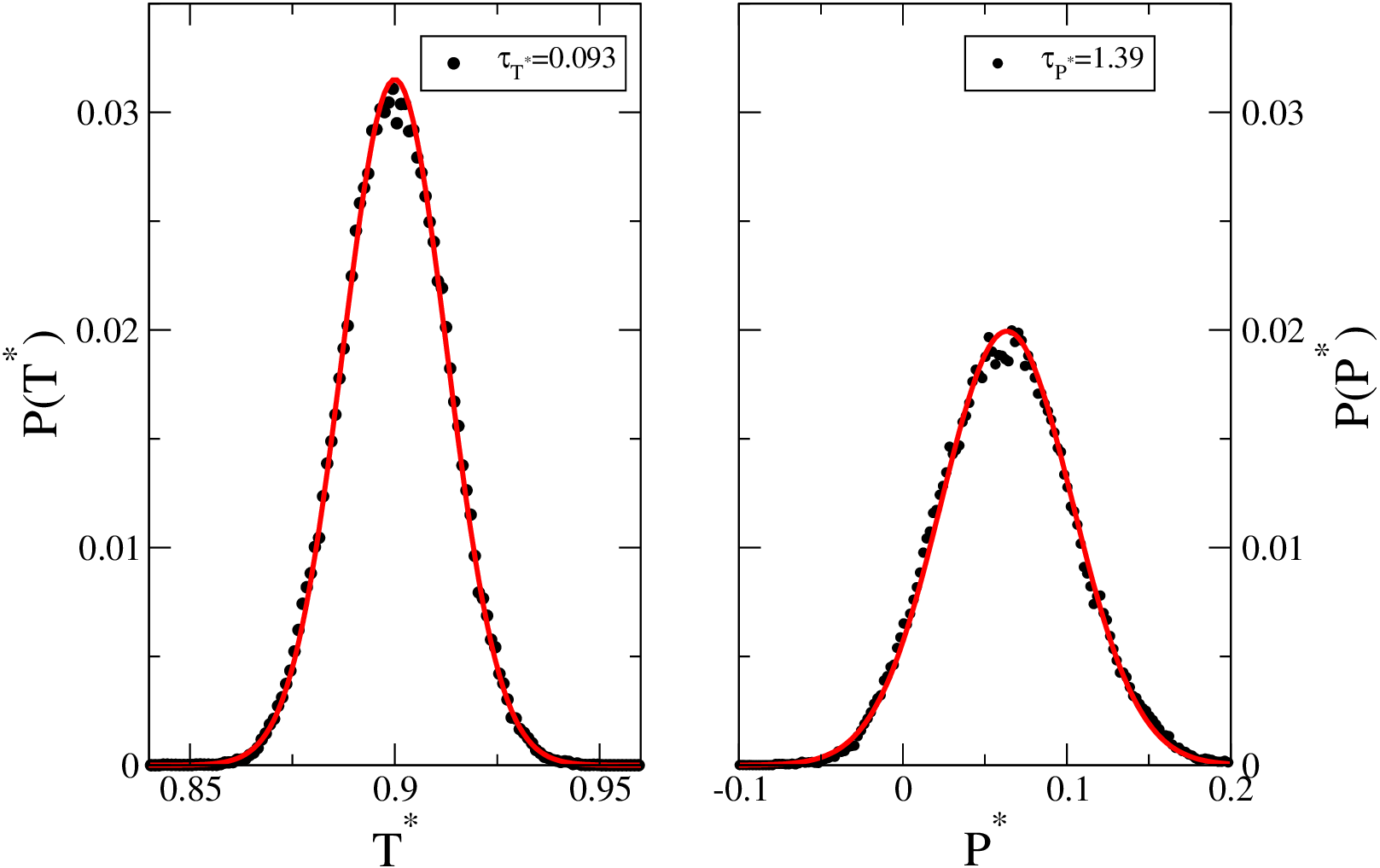}
\end{figure}
\vspace*{1.0cm}
Figure \ref{fig:tempPdis}

\newpage
\vspace*{2cm}

\begin{figure}[h]
\epsfxsize=12.0cm
\epsfbox{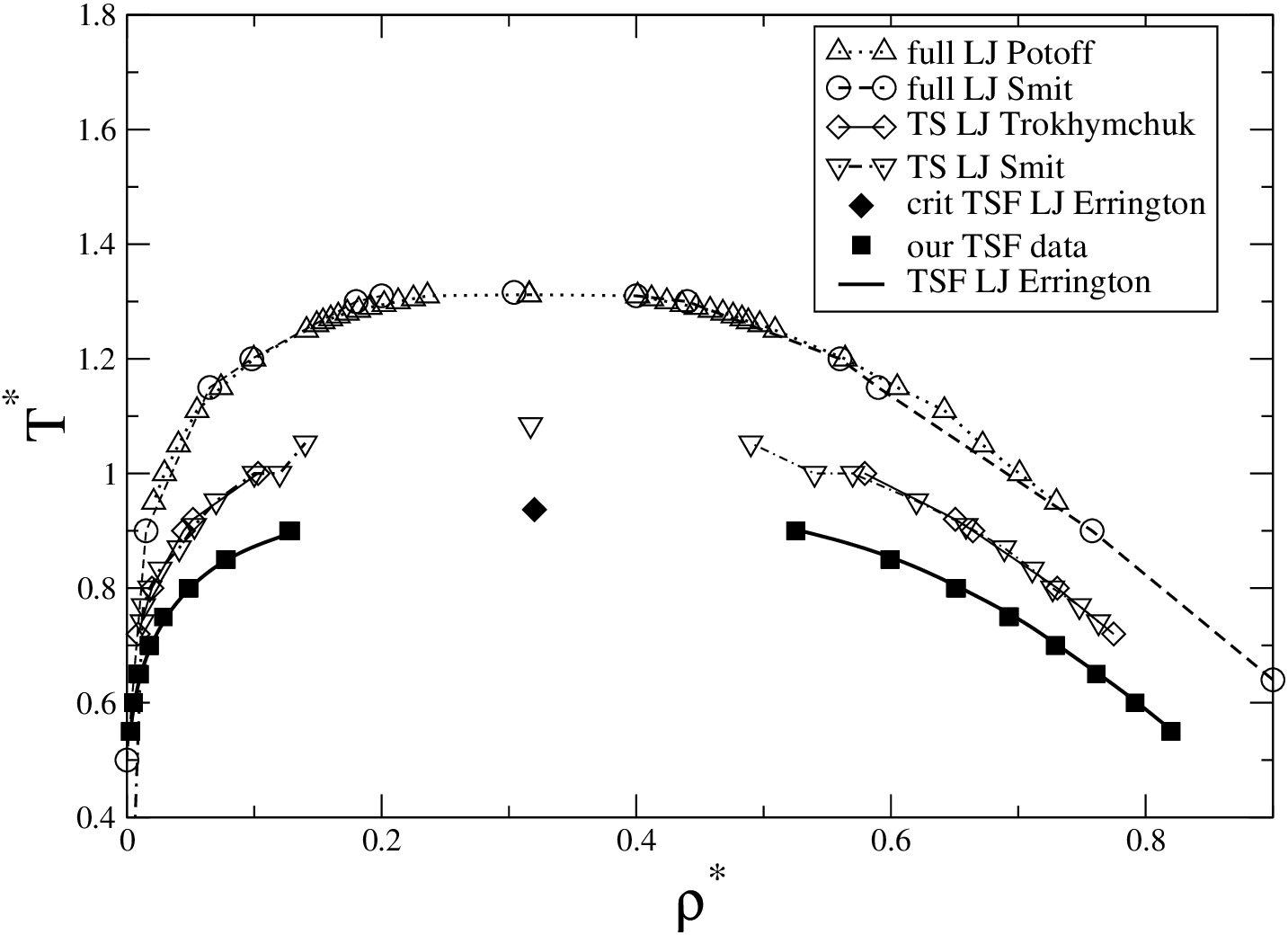}
\end{figure}
\vspace*{1.0cm}
Figure \ref{fig:phasediagrame1}

\newpage
\vspace*{1cm}

\begin{figure}[h]
\epsfxsize=12.0truecm
\epsfbox{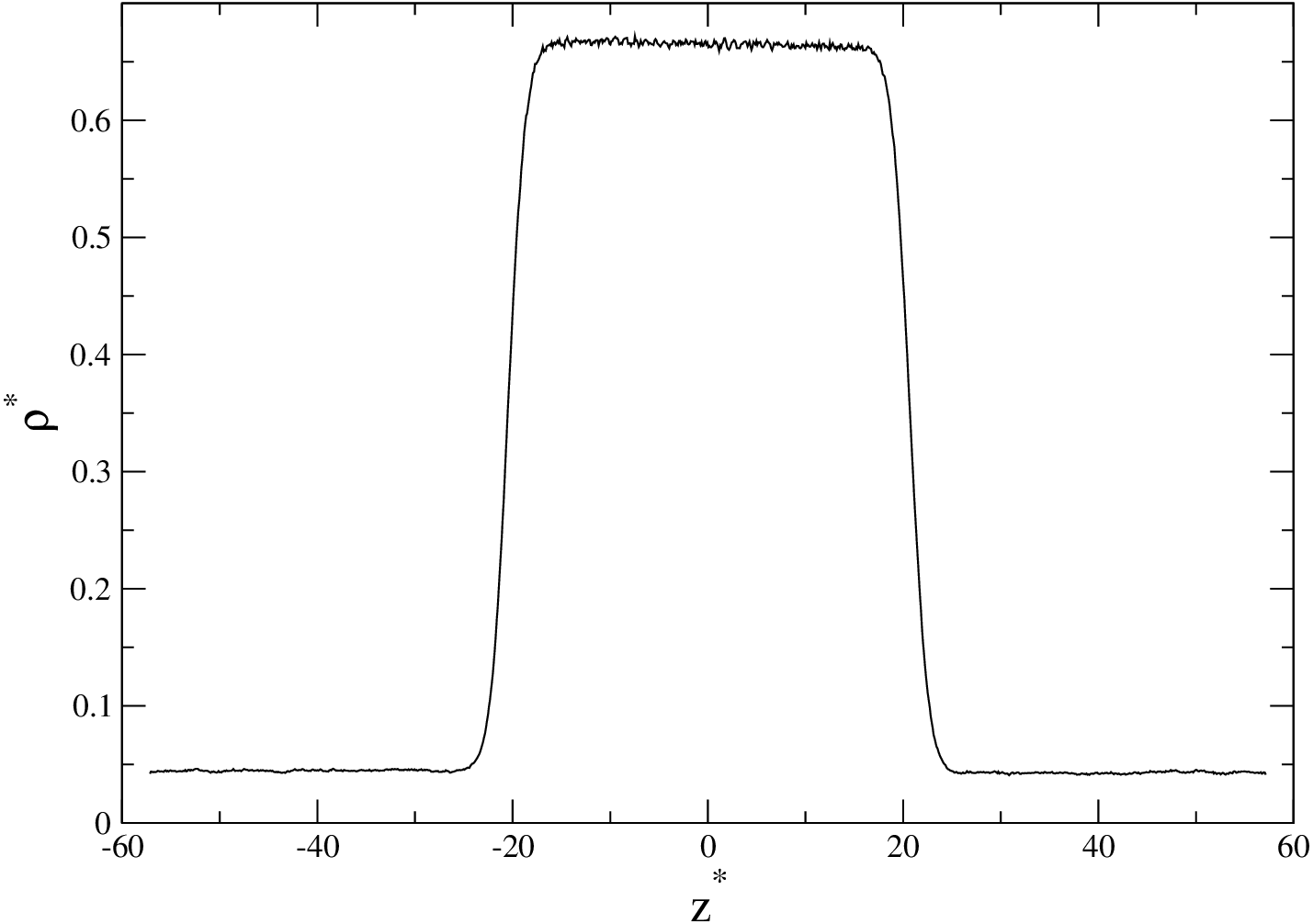}
\vspace*{1.0 cm}

\epsfxsize=12.0truecm
\epsfbox{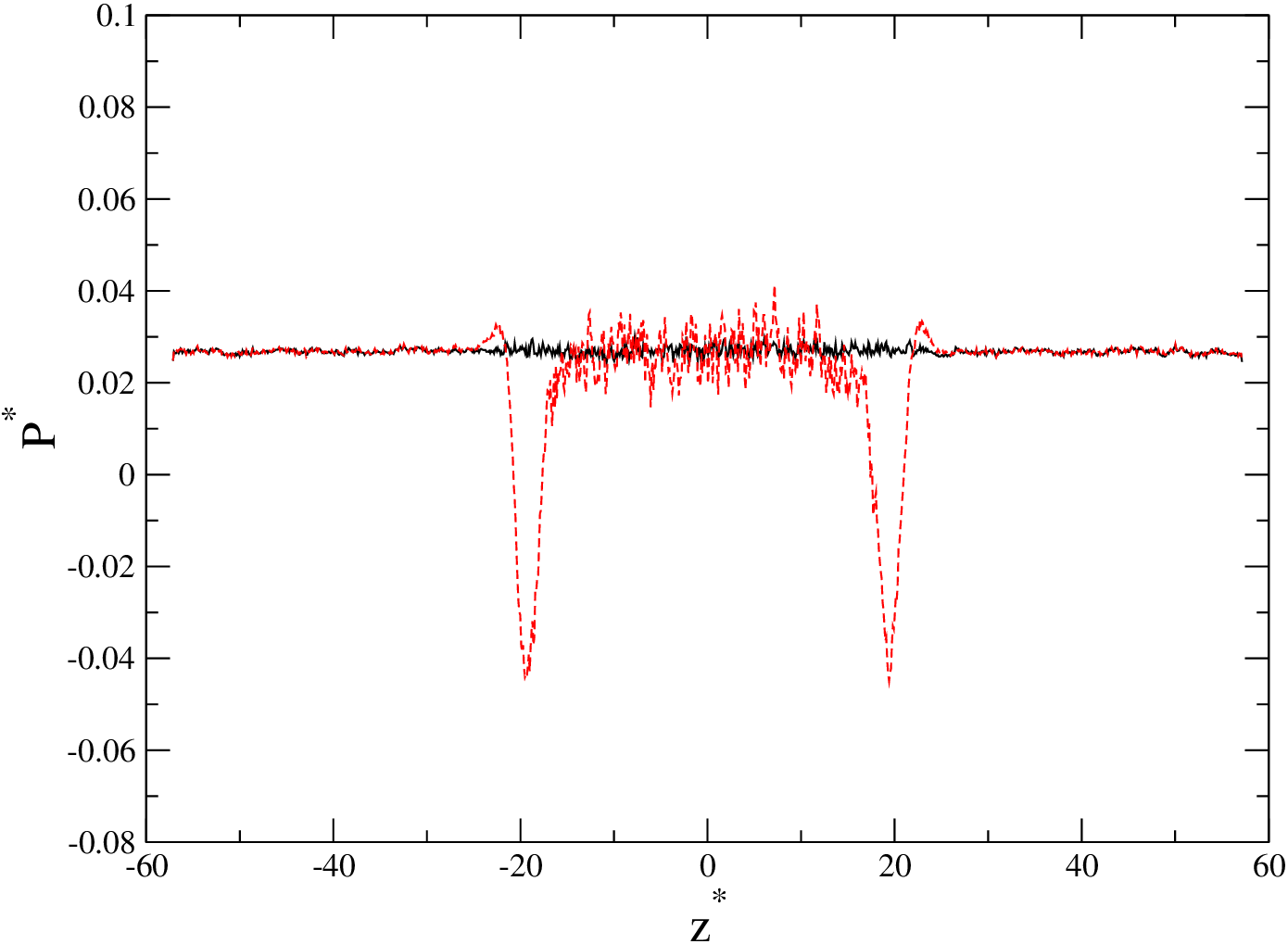}
\end{figure}
\vspace*{1cm}
Figure \ref{fig:profilerhoP}

\end{document}